\documentclass[twocolumn,showpacs,preprintnumbers,amssymb,pra]{revtex4}
\usepackage{graphicx}
\usepackage{dcolumn}
\usepackage{color}
\usepackage{bm}
\begin{document}

\title {\bf Teleportation of two-mode squeezed states}
\author{Satyabrata Adhikari}
\altaffiliation{satyabrata@bose.res.in}
\affiliation{S. N. Bose National Centre for Basic Sciences,
Salt Lake, Kolkata 700 098, India}
\author{A. S. Majumdar}
\altaffiliation{archan@bose.res.in}
\affiliation{S. N. Bose National Centre for Basic Sciences,
Salt Lake, Kolkata 700 098, India}
\author{N. Nayak}
\altaffiliation{nayak@bose.res.in}
\affiliation{S. N. Bose National Centre for Basic Sciences,
Salt Lake, Kolkata 700 098, India}
\date{\today}

\vskip 0.5cm
\begin{abstract}
We consider two-mode squeezed states which are parametrized by the squeezing
parameter and the phase. We present a scheme for teleporting such entangled 
states of continuous
variables from Alice to Bob. Our protocol is operationalized through 
the creation 
of a four-mode entangled state shared by Alice and Bob using linear
amplifiers and beam splitters. Teleportation of the entangled state proceeds
with local operations and the classical communication of four bits.
We compute the fidelity
of teleportation and find that it exhibits a trade-off with
the magnitude of entanglement of the resultant teleported state.
\end{abstract}


\maketitle

\section{Introduction}

Quantum teleportation is an important and vital quantum information processing
task where an arbitrary unknown quantum state can be replicated at a distant
location using previously shared entanglement and classical communication 
between the sender and the receiver. A remarkable application of entangled
states having many ramifications in information technology, quantum 
teleportation can also be combined with other operations to construct advanced
quantum circuits useful for information processing\cite{books}.
The original teleportation protocol of Bennett et al.\cite{bennett} 
for an unknown qubit using an EPR pair has
been generalized to the case of non-maximally entangled or a noisy channel
between the sender and the receiver\cite{noisy}. The loss of fidelity for
teleportation using non-maximally entangled channels could be compensated
by schemes for probabilistic teleportation\cite{probab}. The first 
experimental demonstration of quantum teleportation was reported 
by Bouwmeester et al.\cite{expt1}.

Quantum teleportation is also possible for systems corresponding to infinite 
dimensional Hilbert spaces\cite{contvar,contvar1,charact,gaussian,nongaussian,tan,johnson,teleclone}.
The teleportation process 
for continuous variables was originally formulated in terms of Wigner 
functions\cite{contvar1} and
has also been extended in terms of characteristic functions\cite{charact} 
of the quantum systems
involved. Schemes for obtaining optimal fidelity 
of teleportation using 
Gaussian\cite{gaussian} as well as non-Gaussian\cite{nongaussian} resource 
states have been devised. The first experiment of continuous 
variable teleportation was
performed by Furusawa et al.\cite{expt2}. Since then there have been further
improvements in the fidelity of teleportation obtained in 
experiments\cite{expt3}.  Recently, an experimental characterization of 
continuous variable quantum communication channels has been established by
shared entanglement together with local operations and classical 
communications\cite{expt4}. 

Since quantum entanglement is fragile and is easily destroyed in distribution,
establishing entanglement between quantum systems at distant locations, and
transporting entanglement from one location to another are
rather challenging tasks. Various ingenious methods have been proposed to
accomplish these, such as by  using entanglement swapping 
protocols\cite{swapp},
quantum repeaters by combining operations of swapping  with entanglement 
purification\cite{repeat}, and by continuous measurements\cite{contmeas}.
For continuous variable systems, some protocols
for entanglement swapping\cite{swapcont}, establishing entanglement between
distant stations through teleportation\cite{tan}, testing the efficiency
of teleportation with the aid of a third party\cite{johnson}, and
combining teleportation with cloning\cite{teleclone} have been 
proposed. But no protocol for the explict teleportation of an entangled
continuous variable state exists in the literature, akin to a similar scheme
for discrete variables for teleporting a two-qubit entangled state,
that has been presented recently\cite{yeo}. 

The aim of this work is to propose an explicit scheme for the teleportation
of an unknown two-mode entangled state of continuous 
variables from one party (Alice) to the
other distant party (Bob).  For this purpose we first show how an entangled
state of four modes can be generated and shared by Alice and Bob with the
help of linear amplifiers and beam-spitters. Our protocol for teleportation
can then proceed in the usual way with Alice making measurements on her side 
and communicating their results classically to Bob who in turns makes a local
operation to obtain the teleported entangled two-mode state. The communication
of four bits of information from Alice to Bob is required, similar to the case
of the protocol for teleporting entangled states of two qubits\cite{yeo}.
We compute the entanglement of the teleported state with Bob, and also the
fidelity of teleportation as functions of the squeezing parameters of the
states generated by the source and the teleportation amplifiers. A 
trade-off between the entanglement of the teleported state and the fidelity
of teleportation is observed with respect to the squeezing parameters.

\section{The teleportation protocol}

\begin{figure}[h!]
\begin{center}
\includegraphics[width=8cm]{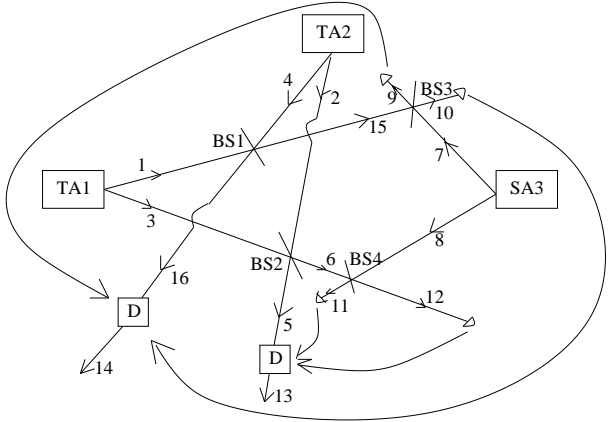}
\caption{Schematic diagram for the teleportation protocol. Alice's amplifier
SA3 generates the two-mode entangled state ($x_7,x_8$) to be teleported. Bob
has two amplifiers TA1 and TA2 and two beam-splitters BS1 and BS2 using which
he generates a four-mode state ($x_5,x_6,x_{15},x_{16}$). He keeps two of
these modes $x_5$ and $x_{16}$ with himself, and sends the remaining two modes
$x_6$ and $x_{15}$ to Alice. Using the beam-splitters BS3 and BS4 Alice
combines her modes $x_7$ and $x_8$ with those sent by Bob, and performs four
measurements on the output modes $x_9,x_{10},x_{11},x_{12}$. She then 
communicates her results to Bob who uses these to apply a unitary 
transformation to displace the modes $x_5$ and $x_{16}$. The final teleported
state is found in the modes $x_{13}$ and $x_{14}$.}
\end{center}
\label{f1}
\end{figure}

Our protocol is as follows.
Let Alice hold the source parametric amplifier SA3 whose two output entangled
modes are to be teleported. Bob possesses two teleportation 
amplifiers TA1 and TA2 which
are required as compulsory accessories for our protocol of teleportation
of two-mode entangled states. Alice's task is to teleport the entangled state
of the modes $x_7$ and $x_8$ as signal and idler 
originating from her SA3 to Bob. But
prior to this one requires to set up a four-mode entangled state to be
shared by Alice and Bob. For this purpose, consider the modes $x_1$ and
$x_3$ coming out of the amplifier TA1, and the modes $x_2$ and $x_4$
coming out of TA2. The covariance matrix of the four modes $x_1,x_2,x_3,x_4$
is given by $\sigma^{(1)(2)(3)(4)} = \sigma^{(1)(3)} \oplus \sigma^{(2)(4)}$,
with $x_i \equiv (X_i,P_i)$, and we assume that the two amplifiers 
are similar, i.e.,
\begin{equation}
\sigma^{(1)(3)} = \sigma^{(2)(4)} = \left(\begin{matrix}{\alpha & \gamma \cr
\gamma^T & \beta}\end{matrix}\right)
\label{covmat1}
\end{equation}
with
\begin{eqnarray}
\alpha &=& \left(\begin{matrix}{c - hs & 0\cr 0 & c+hs}\end{matrix}\right)\nonumber \\
\beta &=& \left(\begin{matrix}{c+hs & 0\cr 0 & c-hs}\end{matrix}\right)\\
\gamma &=& \left(\begin{matrix}{ks & 0\cr 0 & -ks}\end{matrix}\right)\nonumber
\label{defs1}
\end{eqnarray}
where $c = \mathrm{Cosh}(2r), s = \mathrm{Sinh}(2r), k = \mathrm{Sin}(2\phi), 
h = \mathrm{Cos}(2\phi)$ with $r$ being the squeezing parameter and $\phi$ the
amplifier phase. Bob uses two beam-splitters (BS1 and BS2) represented by the
matrix $B_1$ as
\begin{equation}
B_1 = \left(\begin{matrix}{I_2/\sqrt{2} & 0 & 0 & I_2/\sqrt{2} \cr
0 & I_2/\sqrt{2} & I_2/\sqrt{2} & 0 \cr 0 & I_2/\sqrt{2} & -I_2/\sqrt{2} & 0\cr
I_2/\sqrt{2} & 0 & 0 & -I_2/\sqrt{2}}\end{matrix}\right)
\label{beamsplit1}
\end{equation}
where, $I_2$ is a $2\times 2$ identity matrix, to mix the modes
($x_1,x_4$) and ($x_2,x_3$), respectively. The CM corresponding to the four
output modes ($x_5,x_6,x_{15},x_{16}$) of the two beam-splitters are given by
\begin{equation}
\sigma^{(5)(6)(15)(16)} = B_1\sigma^{(1)(3)(2)(4)}B_1^{\dagger}
\label{covmat2}
\end{equation}
Bob then supplies the modes $x_{15}$ and $x_6$ to Alice and keeps the 
remaining modes $x_5$ and $x_{16}$ with himself. Hence, Alice and Bob share
a four-mode entangled state to be used for the teleportation protocol.

Alice has the two entangled modes $x_7$ and $x_8$ originating from her 
source amplifier SA3 to teleport, in addition to the two
modes $x_{15}$ and $x_6$ which Bob has sent her. The combined six-mode
state (four modes with Alice and two with Bob) are represented by the CM
\begin{eqnarray}
\sigma^{(5)(6)(15)(16)(7)(8)} = \sigma^{(5)(6)(15)(16)} \oplus \sigma^{(7)(8)}\nonumber\\
\sigma^{(7)(8)} = \left(\begin{matrix}{x-uy & 0 & vy & 0\cr 0 & x+uy & 0 & -vy\cr vy & 0 & x+uy & 0\cr 0 & -vy & 0 & x-uy}\end{matrix}\right)
\label{covmat3}
\end{eqnarray}
where $x=\mathrm{Cosh}(2q)$, $y=\mathrm{Sinh}(2q)$, $u=\mathrm{Cos}(2\eta)$,
$v=\mathrm{Sin}(2\eta)$, with $q$ being the squeezing parameter of the 
two-mode state ($x_7$ and $x_8$) with Alice, and $\eta$ being the phase of
amplifier SA3.

To proceed further with the teleportation protocol, Alice uses two 
beam-spiltters BS3 and BS4 represented by
\begin{equation}
B_2 = \left(\begin{matrix}{I_2/\sqrt{2} & 0 & 0 & 0 & I_2/\sqrt{2} & 0 \cr
0 & I_2 & 0 & 0 & 0 & 0 \cr 0 & 0 & I_2/\sqrt{2} & 0 & 0 & I_2/\sqrt{2} \cr
0 & 0 & 0 & I_2 & 0 & 0 \cr I_2/\sqrt{2} & 0 & 0 & 0 & -I_2/\sqrt{2} & 0 \cr
0 & 0 & I_2/\sqrt{2} & 0 & 0 & -I_2/\sqrt{2}}\end{matrix}\right)
\label{beamsplit2}
\end{equation}
to combine the modes ($x_{15}$ and $x_7$) and 
($x_6$ and $x_8$), respectively in order to obtain the four output
modes ($x_9,x_{10},x_{11},x_{12}$). Alice then makes measurements on these
four modes. Without loss of generality, let us assume that her choice of
measurements leads to the results ($X_9,P_{10},X_{11},P_{12}$), respectively,
which she communicates to Bob. 

Bob then uses these measurement results to displace the state of the modes
$x_5$ and $x_{16}$ with him, by applying the unitary transformation
\begin{equation}
U = \left(\begin{matrix}{-\sqrt{\frac{2}{3}} & -\sqrt{\frac{1}{3}} & 0 & 0 & 0 & 0 & 0 & 0\cr
0 & 0 & -\sqrt{\frac{1}{3}} & 0 & 0 & 0 & \sqrt{\frac{2}{3}} & 0\cr 0 & 0 & 0 & \sqrt{\frac{2}{3}} & 
\sqrt{\frac{1}{3}} & 0 & 0 & 0\cr 0 & 0 & 0 & 0 & 0 & \sqrt{\frac{1}{3}} & 0 & -\sqrt{\frac{2}{3}}}\end{matrix}\right)
\label{unitary}
\end{equation}
to obtain finally the modes $x_{13}$ and $x_{14}$. The gains $\sqrt{2}$ on the
classical measurements have been chosen such that the resultant matrix 
$\sqrt{3}UKB_2$ (with $K$ defined below) has all the elements 
either $0$, $1$ or $-1$.
The combined physical processes (the beam splitters used by Alice, 
the measurements
performed by Alice, and the unitary transformation performed by Bob) takes
the CM $\sigma^{(5)(6)(15)(16)(7)(8)}$ to
\begin{equation}
\sigma^{(13)(14)} = (UKB_2)\sigma^{(5)(6)(15)(16)(7)(8)}(UKB_2)^{\dagger}
\label{covmat4}
\end{equation}
where
\begin{equation}
K = \left(\begin{matrix}{1 & 0 & 0 & 0 & 0 & 0 & 0 & 0 & 0 & 0 & 0 & 0\cr
0 & 0 & 1 & 0 & 0 & 0 & 0 & 0 & 0 & 0 & 0 & 0 \cr 0 & 0 & 0 & 1 & 0 & 0 & 0
& 0 & 0 & 0 & 0 & 0 \cr 0 & 0 & 0 & 0 & 1 & 0 & 0 & 0 & 0 & 0 & 0 & 0 \cr
0 & 0 & 0 & 0 & 0 & 0 & 1 & 0 & 0 & 0 & 0 & 0 \cr 0 & 0 & 0 & 0 & 0 & 0 & 0 &
1 & 0 & 0 & 0 & 0 \cr 0 & 0 & 0 & 0 & 0 & 0 & 0 & 0 & 0 & 1 & 0 & 0 \cr
0 & 0 & 0 & 0 & 0 & 0 & 0 & 0 & 0 & 0 & 0 & 1}\end{matrix}\right)
\label{measure}
\end{equation}
incorporates the measurements made by Alice. Note that the measurements 
on the four modes $(x_9,x_{10},x_{11},x_{12})$ with the choice of outputs
$(X_9,P_{10},X_{11},P_{12})$ that we have made, corresponds to integrating
out these variables, and hence in the CM (\ref{measure}) the corresponding
rows are deleted.

Thus, the teleported signal and idler can be found as the modes $x_{13}$ and
$x_{14}$ with Bob, represented by the CM
\begin{equation}
\sigma^{(13)(14)} = \left(\begin{matrix}{\sigma_{11} & 0 & \sigma_{13} & 0 \cr
0 & \sigma_{22} & 0 & \sigma_{24} \cr \sigma_{31} & 0 & \sigma_{33} & 0 \cr
0 & \sigma_{42} & 0 & \sigma_{44}}\end{matrix}\right)
\label{covmat5}
\end{equation}
where
\begin{eqnarray}
\sigma_{11} &=& \frac{2c + 2ks + x -uy}{3}\nonumber\\
\sigma_{13} &=& \sigma_{31} = -\frac{vy}{3}\nonumber\\
\sigma_{22} &=& \frac{2c+2ks+x+uy}{3}\\
\sigma_{24} &=& \sigma_{42} = \frac{vy}{3}\nonumber\\
\sigma_{33} &=& \frac{2c+2ks+x+uy}{3}\nonumber\\
\sigma_{44} &=& \frac{2c+2ks+x-uy}{3}\nonumber
\label{defs2}
\end{eqnarray}
The CM $\sigma^{(13)(14)}$ can also be expressed as
\begin{equation}
\sigma^{(13)(14)} = (\sigma')^{(7)(8)} + 2(c+ks)I
\label{covmat6}
\end{equation}
where
\begin{equation}
(\sigma')^{(7)(8)} = \left(\begin{matrix}{x-uy & 0 & -vy & 0\cr
0 & x+uy & 0 & vy\cr -vy & 0 & x+uy & 0\cr 0 & vy & 0 & x-uy}\end{matrix}\right)
\label{covmat7}
\end{equation}
and $I$ is the $4\times 4$ identity matrix. For $k=-1$, and in the limit
$r \rightarrow \infty$, one obtains $\sigma^{(13)(14)} = (\sigma')^{(7)(8)}$. 
It can be shown that the Gaussian state with CM $(\sigma')^{(7)(8)}$ and
the Gaussian state with CM $\sigma^{(7)(8)}$ are equivalent under
local linear unitary Bogoliubov operations (LLUBOs)\cite{llubo}. Thus,
in the limit of ideal input squeezing, the two-mode eintangled
state is teleported perfectly. Further,
if the output states from the amplifiers TA1 and TA2 are coherent states,
i.e., $r=0$, then the variances of the modes $x_7$ and $x_8$ are increased
by twice the level of the vacuum noise. Thus our teleportation protocol
in this special case reproduces the results obtained by Tan\cite{tan} 
for a scheme of teleporting a single-mode state.

\section{Entanglement and fidelity of the output modes}

\begin{figure}[h!]
\begin{center}
\includegraphics[width=9cm]{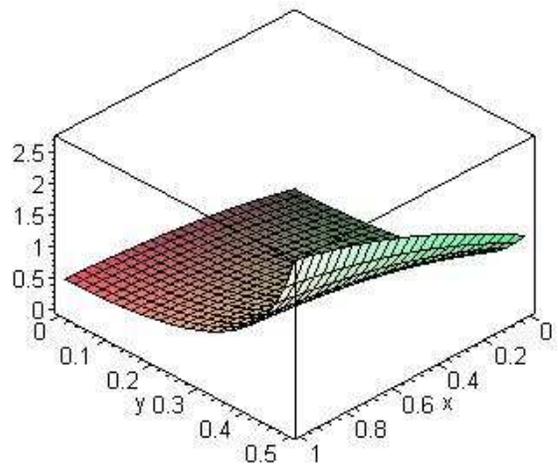}
\caption{(Coloronline) The logarithmic negativity  $E_n$ for the final
teleported state is plotted 
versus the 
squeezing $q$ of the two-mode state generated from the source amplifier SA3 
($x$-axis), and the squeezing $r$
of the two two-mode states from the teleportation 
amplifiers TA1 and TA2, respectively 
($y$-axis).}
\end{center}
\label{f2}
\end{figure}

In order to check that the teleported modes $x_{13}$ and $x_{14}$ indeed
represent an entangled pair, we compute the symplectic eigenvalues of
the partial transpose of the 
CM $\sigma^{(13)(14)}$ given by Eq.(\ref{covmat5}). 
For the modes
$x_{13}$ and $x_{14}$ to be entangled, the smallest symplectic eigenvalue
of $\tilde{\sigma}^{(13)(14)}$ has to be less than one\cite{adesso}, i.e.,
$\tilde{\nu}_{-} < 1$. For simplicity, we assume that the phase of the 
amplifier SA3 is chosen such that $u=0$ and $v=1$, and similarly, the 
phases of TA1 and TA2 are such that $h=k=1/\sqrt{2}$. Then $\tilde{\nu}_{-}$
is given by
\begin{equation}
\tilde{\nu}_{-} = \frac{1}{3}\sqrt{(2c+x-y)^2-2s^2}
\label{eigenvalue}
\end{equation}
The magnitude of entanglement in the teleported state is given by
the logarithmic negativity defined as
\begin{equation}
E_N = \mathrm{max}[0,-\mathrm{log}_2\tilde{\nu}_{-}]
\label{entang}
\end{equation} 
$E_N$ is plotted in Fig.2 versus the squeezing parameters $q$ corresponding
to the two-mode state $(x_7,x_8)$ to be teleported, and $r$ corresponding
to the states $(x_1,x_3)$ and $(x_2,x_4)$ originating from the amplifiers
TA1 and TA2 respectively. One sees that the magnitude of 
entanglement of the teleported
output two-mode state obtained by Bob goes up with increased squeezing of
the states coming from both the source and the teleportation amplifiers.

\begin{figure}[h!]
\begin{center}
\includegraphics[width=9cm]{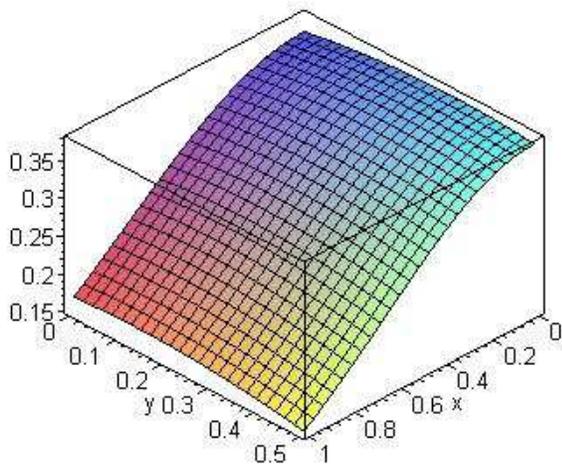}
\caption{(Coloronline) The fidelity of teleportation $F$ is plotted versus the 
squeezing $q$ of the state to be teleported ($x$-axis), and the squeezing $r$
of the output modes of the amplifiers TA1 and TA2 ($y$-axis).}
\end{center}
\label{f3}
\end{figure}

We finally compute the fidelity of the teleported entangled state. The
fidelity can be obtained from the expression\cite{fidelity} given by
\begin{equation}
F = \frac{1}{\sqrt{\mathrm{Det.}[\sigma_{in} + \sigma_{out}]+\delta}-\sqrt{\delta}}
\label{fidel}
\end{equation}
where
\begin{equation}
\delta = 4(\mathrm{Det}[\sigma_{in}] - \frac{1}{4})(\mathrm{Det}[\sigma_{out}] - \frac{1}{4})
\label{defs3}
\end{equation}
with $\sigma_{in}=\sigma^{(7)(8)}$ and $\sigma_{out}=\sigma^{(13)(14)}$ in
the present case given by Eqs.(\ref{covmat3}) and (\ref{covmat5}) respectively.
We plot the fidelity of teleportation versus the squeezing parameters in 
Fig.3. We notice that a maximum fidelity of $0.38$ is possible in this
scheme for an initial two-mode coherent state generated by the source 
amplifier. The fidelity stays nearly constant with variation of the squeezing
of the states from the teleportation amplifiers. Thus coherent states generated
by the teleportation amplifiers can also be used to implement this protocol.
However, increased squeezing of the initial state from the source amplifier
leads to the loss of fidelity. The latter feature is in sharp contrast to the
magnitude of entanglement of the final two-mode state with Bob, which increases
with the increase of squeezing of Alice's two-mode state. 
This result seems
to support the contention of Johnson et al.\cite{johnson} that the average
fidelity may not be a good indicator of the quality of quantum teleportation.
Though their results were obtained in the context of teleportation of single
modes together with the physical transport of modes between three 
parties\cite{johnson},
it might be more appropriate to use the measure of `entanglement 
fidelity'\cite{contvar1,entfidel} to quantify, where possible, the ability 
of a process to preserve entanglement.

\section{Conclusions}

To summarize, we have presented the first explicit protocol for  
teleportation of two-mode entangled squeezed states. Our scheme is 
accomplished by the creation of a four-mode state 
shared initially by two distant parties through beams generated by two 
teleportation amplifiers and combined by two beam splitters. Teleportation 
takes place with the usual local measurements, unitary operations and the
classical communication of four bits parallel to the case involving
entangled states of discrete variables\cite{yeo}. If coherent states from
teleportation amplifiers are used to create the four-mode
state shared by the two parties for enabling teleportation, the variance
of the entangled two-mode teleported state increases by twice the level of
the vacuum noise. In general, the entanglement 
obtained for the teleported
state increases with the squeezing of the initial two-mode state which
however leads to loss of fidelity of teleportation. It might be worthwhile
to explore other schemes of teleportation to check whether fidelity
of teleported entangled states could be substantially incresed.

\end{document}